\documentclass[letterpaper, 10 pt, journal, twoside]{ieeetran}

\IEEEoverridecommandlockouts                              

\usepackage{amsmath} 
\usepackage{amssymb}  
\usepackage{epsfig} 
\usepackage{amsmath} 

\usepackage{epstopdf}
\usepackage{epstopdf}
\usepackage{algorithm}
\usepackage{algpseudocode}
\usepackage{color}
\usepackage{xcolor}

\usepackage{url}
 \usepackage{subfigure}

\usepackage[colorinlistoftodos]{todonotes}

\usepackage{graphicx}
\usepackage[version=4]{mhchem}
\usepackage{siunitx}
\usepackage{longtable,tabularx}
\usepackage{varioref}
 \usepackage{wrapfig}
 \usepackage{threeparttable}
 \usepackage{dcolumn}
  \newcolumntype{d}{D{.}{.}{-1}}
 \usepackage{nomencl}
  \makeglossary
 \usepackage{subfigure}
 \usepackage{subfigmat}
 \usepackage{fancyvrb}
  \fvset{fontsize=\footnotesize,xleftmargin=2em}
 \usepackage{lettrine}
\usepackage{epstopdf}
\usepackage{graphicx}
\graphicspath{{./figure/}}
\usepackage{amsfonts}
\usepackage{color}

\newcommand{\beq}{\begin{equation}}
\newcommand{\eeq}{\end{equation}}
\newcommand{\beqs}{\begin{equation*}}
\newcommand{\eeqs}{\end{equation*}}
\newcommand{\bea}{\begin{eqnarray}}
\newcommand{\eea}{\end{eqnarray}}
\newcommand{\beas}{\begin{eqnarray*}}
\newcommand{\eeas}{\end{eqnarray*}}
\newcommand{\sgn}{\text{sgn}}

\title{Sliding Mode Control Techniques and Artificial Potential Field for
Dynamic Collision Avoidance in Rendezvous Maneuvers}

\author{Mauro Mancini$^{1}$, Nicoletta Bloise $^{1}$, Elisa Capello$^{2}$ and Elisabetta Punta$^{3}$%
\thanks{$^{1}$M. Mancini and N. Bloise are with the Department
of Mechanical and Aerospace Engineering, Politecnico di Torino, Corso Duca degli Abruzzi 24, 10129 Torino, Italy,}%
\thanks{$^{2}$E. Capello is with the Department of Mechanical and Aerospace Engineering, Politecnico di Torino and with the CNR-IEIIT, Politecnico di Torino, Corso Duca degli Abruzzi 24, 10129 Torino, Italy}%
\thanks{$^{3}$ E. Punta is with the CNR--IEIIT, Politecnico di Torino, Corso Duca degli Abruzzi 24, 10129 Torino, Italy.}%
}

\begin{document}

\maketitle
\thispagestyle{empty}
\pagestyle{empty}

\begin{abstract}
The paper considers autonomous rendezvous maneuver and proximity operations of two spacecraft in presence of obstacles. A strategy that combines guidance and control algorithms is analyzed. 
The proposed closed-loop system is able to guarantee a safe path in a real environment, as well as robustness with respect to external disturbances and dynamic obstacles. 
The guidance strategy exploits a suitably designed Artificial Potential Field (APF), while the controller relies on Sliding Mode Control (SMC), for both position and attitude tracking of the spacecraft. 
As for the position control, two different first order SMC methods are considered, namely the component-wise and the simplex-based control techniques. 
The proposed integrated guidance and control strategy is validated by extensive simulations performed with a six degree-of-freedom (DOF) orbital simulator and appears suitable for real-time control with minimal on-board computational effort. Fuel consumption and control effort are evaluated, including different update frequencies of the closed-loop software.
\end{abstract}


\section{Introduction} 
\label{sec:intro}

In the aerospace field, the rendezvous maneuver consists in a series of maneuvers, usually divided into subphases~\cite{fehse2003automated}, which leads to perform a dock between two spacecraft, a passive Target and an active Chaser. Since this maneuver is essential for space missions and explorations,
several studies and missions have been performed to realize it autonomously. The on-board software usually includes collision-free path planning, to avoid impacts with other spacecraft but also with countless space debris, thanks to a detection system. 
In~\cite{luo2014survey} Luo and co-authors propose a survey of orbital dynamics and control, in which the related equations and a brief description of some automated control methods are presented, not including Guidance, Navigation and Control (GNC) algorithms. 
Autonomous operations require online robot navigation and obstacle avoidance strategies. APF methods shape the environment with artificial harmonic potentials,~\cite{Kha86}. These strategies plan the trajectories of the robot along the gradient of the APF, which represents the environment and is generated placing positive charges in the obstacles and a negative charge in the Target.
Control strategies must ensure that the robot tracks the APF gradient, despite uncertainties and disturbances.

Since the rendezvous maneuver is a nonlinear maneuver with regard to dynamics and kinematics, position and attitude control must ensure high accuracy and excellent robustness against external disturbances and parameter variations with simple design.
SMC,~\cite{utkin1992sliding,Fridman_book}, are nonlinear control techniques with remarkable properties of precision, robustness and ease of design, tuning, and implementation.
SMC design consists of two fundamental steps. First a sliding surface is defined such that if the trajectories of the system belong to that surface the closed loop system is stable and performs as expected.
Afterwards, the control law is selected, which is able to drive and keep the system on the sliding surface, i.e. to enforce a sliding motion on the surface.
The closed loop response of the system in sliding mode is insensitive to uncertainties of various types, namely uncertainties of the model parameters, disturbances and nonlinearities, under certain boundedness and matching conditions.
%
%
From a practical point of view, SMC techniques allows to control nonlinear processes subject to external disturbances and large model uncertainties.
The great effectiveness of SMC to control nonlinear uncertain systems is obtained in spite of relatively simple design of the controllers. SMC can guarantee accuracy and robustness at the same time. One of the main feature of SMC is the exploitation of discontinuous control laws (first order SMC) or instead of continuous control inputs obtained with discontinuous terms in the time derivatives of the control inputs (second order SMC). The simplex-based SMC,~\cite{BajIzo85,Bart1997,Bart2004,BarPun15}, is a method of the first order, which is particularly well suited to be adopted when the actuation system consists of mono-directional discontinuous devices,~\cite{BarCocPun00,capello2017simplex}. 

%
Guldner and Utkin,~\cite{guldner1995sliding}, used a combination of APF and SMC for motion planning in the robotic field. In recent years, some papers have proposed APF for guidance algorithm for proximity operations. In ~\cite{zappulla2016experiments} an adaptive APF is described and its computational efficiency is demonstrated by the experimental testbed. The problem of obstacle constraints is also addressed in~\cite{zappulla2016experiments} via an APF method and an optimal SMC. 


As discussed in~\cite{utkin1992sliding} and analyzed for space maneuvers in~\cite{capello2017sliding}, internal and external disturbances acting on the system shall be dealt with for real implementation.
For the considered spacecraft model, a first order SMC method appears to be the most appropriate because the system is actuated by discontinuous mono-directional thrusters that can be only switched on and off.
In this paper, two first order SMC are proposed for the position control, namely simplex-based and component-wise SMC, which are compared in simulation.
The simplex-based SMC system is proposed to reduce the required number of thrusters to fully actuate the system. Simulations show that, reducing the commutations among the allowed control structures, the simplex-based SMC reduces the fuel consumption necessary to perform the desired maneuver. 
The attitude control is carried out with the second order super-twisting (STW) SMC~\cite{levant1993sliding}, which can be suitably implemented by the continuous actuation system of the spacecraft, provided by reactions wheels. 

A combination of this guidance and control method for rendezvous operations is treated in~\cite{bloise2017obstacle}, not considering dynamic obstacles.
Some interesting novelties in this paper are introduced: i) a path planning evaluated with moving obstacles (dynamic obstacles), 
ii) a sensor is included for the detection of the obstacles, 
iii) two different first order SMC for position control are compared in order to find how to reduce the control effort, and iv) hardware and software constraints are taken into account in terms of variation of the update (switching) frequency of the closed-loop system with the combined algorithms.
The paper is organized as follows. In Section~\ref{sec:Mathematical Models} the models of spacecraft and actuators are introduced. In Section~\ref{sec:Guidance Strategy} the guidance algorithms based on APF are proposed and the dynamic collision avoidance is detailed. The position control problem of the spacecraft is considered in Section~\ref{sec:Control Strategy}; two different first order SMC are proposed, namely a component-wise SMC and a simplex-based one. The attitude controller is briefly introduced. Simulation results are presented in Section~\ref{sec:Simulation Results}. Conclusions are drawn in Section~\ref{sec:Conclusions}.

\section{Mathematical Models}
\label{sec:Mathematical Models}
The developed orbital simulator includes six degree-of-freedom (DOF) spacecraft dynamics, the actuation system model and other mathematical models, such as model of sensors errors and of external disturbances. Furthermore, actuator errors and nonlinearities are also taken into account. For the details on the spacecraft dynamics refer to \cite{fehse2003automated,markley2014fundamentals,capello2017sliding}.

\subsection{Rendezvous Maneuver}
In our case study, the last two phases of the rendezvous maneuver are analyzed, starting when the Chaser has already reached the Target orbit. 
For the position dynamics, a Local-Vertical-Local-Horizontal (LVLH) frame is employed, where V-bar is directed along the orbital speed, R-bar along the direction from the origin of this frame to the center of the Earth and H-bar along the opposite direction of the orbital angular momentum. 
Therefore, a more complete description can be found in~\cite{fehse2003automated}, where the \textit{radial boost} and the \textit{cone approach} are explained in detail. 


\subsection{Spacecraft Dynamics}
\label{Spacecraft Dynamics}

The spacecraft dynamics includes both position and attitude dynamics. Since the maneuver starts few kilometers far from the Target, the Hill equations are used for the definition of the orbital dynamics. This set of simplified equations can be used for distances between Chaser and Target vehicles that are very small compared with the distance to the centre of the Earth, without affecting the effectiveness of the proposed approach. Moreover, a circular orbit is considered.
These equations are written with respect to the origin of a LVLH frame, centered on the Target satellite~\cite{fehse2003automated}. 
So, the position dynamics can be expressed as
\bea
\label{Hill}
\begin{cases}
\begin{aligned}
\ddot{ x} &= \frac{F_x}{m_c} + 2\omega_0\dot{z}, \\
\ddot{y} &= \frac{F_y}{m_c} -\omega^2_0 y, \\
\ddot{z} &= \frac{F_z}{m_c} - 2\omega_0\dot{ x}+3\omega_0^2 z ,
\end{aligned} 
\end{cases}
\eea 

where $\ddot{x}$, $\ddot{y}$ and $\ddot{z}$ are accelerations, $\dot{x}$, $\dot{y}$ and $\dot{z}$ are velocities, $x$, $y$, and $z$ are Hill positions, all the variables are written in LVLH frame. 
The Chaser mass $m_c$ varies with time as follows $\frac{|F|}{g_0I_{SP}}$, with $|F| = |F_x|+|F_y|+|F_z| \in \mathbb{R}$ the total force acting on the vehicle, $g_0 \in \mathbb{R}$ the constant of gravity
at sea level and $I_{SP} \in \mathbb{R}$ the thruster specific impulse, $\omega_0$ is the orbital angular velocity of the reference LVLH frame (centered in the Target).
$F = [F_x, F_y, F_z]^T \in \mathbb{R}^3$ is the total force acting on the spacecraft and it includes both forces due to the thrusters system and external disturbances, so $F = F_{\mathrm{thr}}+\Delta F_{ex} \in \mathbb{R}^3$, in which $F_{\mathrm{thr}} \in \mathbb{R}^3$ is the force provided by the thrusters (i.e. by the control system) and $\Delta F_{ex} \in \mathbb{R}^3$ is due to the external disturbances. 
These latter are estimated as functions of the orbital altitude and the size of the satellite \cite{markley2014fundamentals}, \cite{sidi1997spacecraft}, and consist of three contributions: (i) the constant aerodynamic drag force, which acts only along V-bar, (ii) the force due to oblateness of the Earth (usually called \textit{J2} effect), and (iii) the force due the solar radiation pressure. 
External forces also give raise to moments 
with two contributions, which are respectively due to: (i) the solar radiation pressure and (ii) the gravity effects.  

If we focus on the simulations scenario, the disturbances, previously described, are included in the vector $\Delta F_{ex}$. In detail, $\Delta F_{ex} = F_{AD}+ F_{J2}+F_{s}$. Then, $\Delta F_{ex}$ is added to the force exerted by the thrusters $F_{\mathrm{thr}}$. Finally, the total force $F = F_{\mathrm{thr}}+\Delta F_{ex}$ is rotated by means a rotation matrix from body frame to LVLH frame in order to insert it in the Hill equations for the position dynamic. See Appendix A for details.

Furthermore, errors in shooting directions and thrusts magnitude of thrusters are included \cite{Wilsona, Wilsonb}.

To include these forces in Equation (\ref{Hill}), a rotation is required since the orientation of the thrusters is known in body frame. Hence, the rotation have to be applied with the matrix $R_{LVLH}$ computed by means Euler angles,  
\beq
\label{Fthr}
F_{\mathrm{thr}} = R_{LVLH}(\phi,\theta,\psi) F^{b}_{\mathrm{thr}} ,
\eeq
where $F^{b}_{\mathrm{thr}} \in \mathbb{R}^3$ is the thrust force expressed in body frame (related to the actuation system) and $R_{LVLHb}(\phi,\theta,\psi)$ is the rotation matrix from body frame to LVLH frame (the order of rotation is fixed and it is 3-2-1)~\cite{markley2014fundamentals}. The attitude angles are evaluated from the quaternion dynamics, as explained in Chapter 3 of \cite{markley2014fundamentals}.
In a similar way, the external disturbances $\Delta F_{ex}$ are
\beqs
\Delta F_{ex} = R_{LVLH}(\phi,\theta,\psi) \Delta F_{ex}^{b} .
\eeqs

The focus of this paper is the position tracking. However, Euler equations and quaternion dynamics (see Chapter 3 and 6 of \cite{markley2014fundamentals}) are modeled in the simulator, to analyze a complete 6 DOF spacecraft dynamics.
In particular, 
the total torque applied to the Chaser consists of three terms: (a) the moment due to the thrusters $M_{thr}$, (b) the torque provided by the reaction wheels (i.e. the actuators devoted to attitude control) $M_{RW}$ and conceived as a STW SMC (see Section \ref{sec:Control Strategy}), and (c) the moment due to external disturbances $\Delta M_{ex}$ (see \cite{markley2014fundamentals} and Appendix A). 
$\Delta M_{ex}$ is composed by two contributions: i) a constant moment of about $10^{-5}~N$ due to the solar pressure and ii) a term of the order of about $10^{-4}~Nm$ produced by the gravitational effects.

\section{Guidance Strategy}
\label{sec:Guidance Strategy}

A guidance algorithm based on the theory of APF is proposed to find the path that leads the Chaser toward the Target, avoiding obstacles. One of the most valuable feature of this method is the ability to update in real-time the planned path. The speed field is evaluated using data concerning the state vector of the spacecraft, the environment in which it moves and the desired final position. Therefore, the low computational effort of APF method  represents a great advantage to achieve an autonomous maneuver and can be easily implemented \textit{on-board}.
This strategy assigns to the desired final point an attractive potential field and a repulsive one related to each obstacle. 

A Light Detection and Ranging (LIDAR) sensor for the obstacle detection is used to determine the distance of an object by using a laser pulse. The selected sensor has a range of about $300$ m and helps to drive the Chaser toward the Target during the last phase of the maneuver in order to guarantee docking requirements.

\subsection{Attractive Potential Field}
\label{sec:Attractive potential field}

As in~\cite{bloise2017obstacle}, the Chaser approaches the goal thanks to a paraboloid attractive artificial potential field defined as 
\begin{equation}
    U_{a}(x)=\frac{1}{2}k_{a}||e(x)||^2 ,
    \label{eq:Ua}
\end{equation}
where $k_a$ is the proportional positive gain,
while $e(x)=x_d-x \in \mathbb{R}^3$ represents the error, i.e. the difference between the desired position and the current one. The components of the position vector are expressed in LVLH reference frame. The relative attractive force is obtained through the gradient of Eq. (\ref{eq:Ua}), as follows
\begin{equation}
    {F}_a(x)=\nabla U_{a}(x)=k_{a}e(x) .
    \label{eq:Fa}
\end{equation}


\subsection{Repulsive Potential Field}
\label{sec:Repulsive potential field}

\begin{figure} 
\centering
\vspace{4mm}
\includegraphics[width=0.35\textwidth]{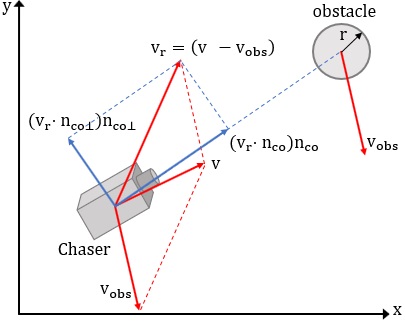}
\caption{Vectors for defining the new repulsive potential}
\label{fig:vel}
\end{figure}

In order to ensure collision avoidance with obstacles, an hyperbolic potential field is constructed around each obstacle ($i=1,\dots\textcolor{blue}{,} N_{obs}$, with $N_{obs}$ number of obstacles). 
\begin{equation*}\begin{split}
\small
    U_{rep,i}(x,v)=
    \begin{cases}
    \small
   \frac{k_{r,i}}{2}&\left(\frac{1}{\eta _i(x)}-\frac{1}{R_{dyn,i}(x,v)}\right)^{2} \\
   &\text{if}\quad\eta_i(x)<\eta_{0,i}\quad\text{and}\quad v_{r,i}\cdot n_{co,i}>0 \\
   0&\text{otherwise}
   \end{cases} ,
   \label{eq:Ur}
\end{split}\end{equation*}
where $k_{r,i}$ is the positive repulsive gain related to the each obstacle, 
$\eta _i(x)=||x-x_{obs, i}||$ is the distance between the Chaser and the \textit{i}-th obstacle,
$v_{r,i}\cdot n_{co,i}$ is the relative velocity between the Chaser and the \textit{i}-th obstacle positive from the Chaser to the obstacle (Fig.~\ref{fig:vel}), $n_{co,i} = \frac{(x_{obs}-x)}{\vert \vert x_{obs}-x\vert \vert_2}$ is the unit vector positive from the Chaser to the \textit{i}-th obstacle,
and $\eta_{0,i}$ is the distance of influence of the obstacle \textit{i}, i.e. maximum $300~m$, based on the used sensor.
In order to take into account both position and speed of the obstacle in the repulsive potential field, the dynamic radius $R_{dyn}$ is introduced instead of $\eta_{0,i}$ \cite{guldner1995sliding, ge2002dynamic}. In this way, a smooth trajectory is obtained and the Chaser velocity is also included.
So,
\begin{equation*}
    R_{dyn,i}(x,v)=\eta_{0,i}+\frac{(v_{r,i}\cdot n_{co,i})^2}{2a_{max}} ,
    \label{R_dyn}
\end{equation*}
in which $a_{max}=\frac{u_x-f}{\sqrt{2}m_c}$ is the acceleration command, with $u_x$ output provided by the thrusters, $m_c$ the mass of the Chaser, and $f>0$ is the constant bound available to the controller, which takes into account the uncertain reduction of the real thrusts provided by the thrusters with respect to the nominal ones.
$U_{rep,i}$ is a function of both relative position and speed.
The repulsive force is given by the negative of the gradient of $U_{rep,i}$ with respect to $x$ and $v$, as follows
\begin{equation}
F_{rep,i}(x,v)=-\nabla_x U_{rep,i}(x,v) - \nabla_v U_{rep,i}(x,v) ,
\label{eq:F_rep}
\end{equation}

where the two gradients are
\begin{equation*}
\begin{split}
    &\nabla_x U_{rep,i}=
    \left(\frac{1}{\eta_i}-\frac{1}{R_{dyn,i}}\right)
    \left( 
    \frac{k_{r,i}}{R^2_{dyn,i}} \nabla_x R_{dyn,i}
    -\frac{k_{r,i}}{\eta_i^2} \nabla_x \eta_i
    \right)
    \\
    &\nabla_v U_{rep,i}=
    \left(\frac{1}{\eta_i}-\frac{1}{R_{dyn,i}}\right) \frac{k_{r,i}}{R^2_{dyn,i}} \nabla_v R_{dyn,i} ,
    \label{eq:F_rep_2}
\end{split}\end{equation*}


in which 
the gradients of the dynamic radius are given by
\begin{equation*}\begin{split}
    \nabla_x R_{dyn,i} = 
    \frac{v_{r,i}\cdot{n}_{co,i}}{a_{max}}\nabla_x\left(v_{r,i}\cdot{n}_{co,i} \right) ,\\
    \nabla_v R_{dyn,i}=
    \frac{v_{r,i}\cdot{n}_{co,i}}{a_{max}}\nabla_v\left(v_{r,i}\cdot{n}_{co,i} \right) ,
    \label{eq:grad_vr} 
\end{split}\end{equation*}
where, according to \cite{ge2002dynamic}, 
\begin{equation*}\begin{split}
    &\nabla_x(v_{r,i}\cdot n_{co,i})=-\frac{1}{\eta_i}\left({v}-{v}_{obs,i}-(v_{r,i}\cdot{n}_{co,i}){n}_{co,i} \right) ,\\
    &\nabla_v(v_{r,i}\cdot n_{co,i})={n}_{co,i},
\end{split}\end{equation*}

and the speed of the obstacle is evaluated as ${v}_{obs}=\tfrac{x_{obs}(t+1)-x_{obs}(t)}{\Delta t_{sensor}}$, with $\Delta t_{sensor}$ sample time of the LIDAR sensor.

The repulsive force (\ref{eq:F_rep}) has two components, ${F}_{rep,i}={F}_{rep~1,i}+{F}_{rep~2,i}$:
\begin{enumerate}
    \item ${F}_{rep~1,i}$ is directed along the line joining Chaser-obstacle; 
    ${F}_{rep~1,i}$ reduces the value of the relative speed $v_{r,i}\cdot n_{co,i}$.
    \item The other component ${F}_{rep~2,i}$ is perpendicular to ${F}_{rep~1,i}$; 
    ${F}_{rep~2,i}$ increases the 
    relative velocity in 
    this perpendicular direction, acting as a steering force for detouring. 
\end{enumerate}

See Fig. \ref{fig:vel} for the definition of all the distances.

\subsection{Total Potential Field}
\label{sec:Total potential field}
The total force resulting from the APF is the sum of the attractive and repulsive forces (\ref{eq:Fa}) and (\ref{eq:F_rep})
\begin{equation}
\small
    {F}_{tot}=
    {F}_a+ \sum_{i=1}^{N_{obs}} {F}_{rep~1,i}+ \sum_{i=1}^{N_{obs}} {F}_{rep~2,i} .
    \label{eq:f_tot}
\end{equation}
%
The force (\ref{eq:f_tot}) is normalized to obtain the direction that the Chaser must follow
\begin{equation}
    E_U=\frac{F_{tot}}{||F_{tot}||} .
    \label{eq:EU}
\end{equation}
The desired velocity vector to be tracked by the Chaser is defined as
\begin{equation}
    \Dot{x}_d=\Dot{x}_{max} E_U , \in \mathbb{R}^3 ,
    \label{eq:v_des}
\end{equation}
where the direction of the Chaser $ E_U $ is given by (\ref{eq:EU}) and 
$\Dot{x}_{max}$ is defined based on specifications and performances required.


\section{Control Strategies}
\label{sec:Control Strategy}
The position and attitude controllers are designed to deal with parametric uncertainties (inaccuracies in the spacecraft model) and bounded uncertainties related to the actuation system, as well as external disturbances.


The spacecraft attitude controller is a SMC STW, the details of which can be found in~\cite{capello2017sliding} and the gains are defined based on the dynamics of the system, as explained in Chapter 6 of \cite{Fridman_book}.

Instead, for the position control two different first order SMC strategies are presented, i.e. a simplex SMC and component-wise SMC. The implementation in real applications of discontinuous control inputs (first order SMC) must take into account the constraints on the switching frequency of the actuators.
Simplex SMC methods reduce the number of thrusters required and implicitly design better shooting strategies.
The results are reduction in fuel consumption and chattering obtained with lower switching frequencies of the control devices.
The effectiveness of the proposed simplex SMC is compared with the results of the component-wise SMC.

The position tracking control problem is considered for the uncertain nonlinear system (\ref{Hill}) actuated by mono-directional devices. The control objective is to steer to zero the suitably chosen sliding output $ \sigma_x \in \mathbb{R}^3$, given by 
\beq
 \sigma_x =  c_x(\dot{x} - {\dot x}_d)  ,
\label{surf_x}
\eeq
where $c_x>0$ is a constant value and ${\dot x}_d = \Dot{x}_{max}E_U$ is the vector of the desired speed given by (\ref{eq:v_des}).


\subsection{Spacecraft Position Control: Simplex SMC}
\label{sec:Simplex}

Let us briefly recall the definition and main features of a simplex of vectors, which can be suitably exploited to design SMC, \cite{Bart1997,Bart2004}.

By definition $ m+1 $ vectors $ v_{i} \in \mathbb{R}^m $, $ i= 1, \ldots\textcolor{blue}{,} m+1 $, form a {\bf simplex of vectors in the $m$-dimensional space $ \mathbb{R}^{m}$} if they generate a convex hull which contains the origin of $ \mathbb{R}^{m} $. Thus there
exist $ m + 1 $ coefficients $ \mu_i > 0 $, $ i= 1, \ldots\textcolor{blue}{,} m+1 $, such that
%
$\sum_{i = 1}^{m+1} \mu_{i} v_{i} = 0$ 
and 
$\sum_{i = 1}^{m+1} \mu_{i} = 1 $.
The vectors $ v_i \in \mathbb{R}^m $, $ i = 1, \ldots, m+1 $, which form a simplex of vectors in $ \mathbb{R}^m $ divide the space $ \mathbb{R}^m $ in $ m + 1 $ non overlapping cones
\begin{equation}
\label{eqt:Qj}
\overline{Q}_j = \mathrm{cone} \left( v_i: \; i = 1, \ldots , {m+1}, \; i \neq j \right) .
\end{equation}
For any point $ x \in \mathbb{R}^m $, there exists an index (or more indices if $x$ belongs to edges) 
$ j \in \{ 1, \ldots , {m+1} \} $ such that
$ x \in \overline{Q}_j $, that is
%
$x = \sum_{i = 1, \, i \neq j}^{m+1} \lambda_i v_i $,
$\lambda_i \geq 0 $ (edges correspond to the case $\lambda_i=0$ for some $i \in  \{ 1, \ldots , {m+1} \} $).
%


In order to design a simplex SMC, let us assume that the spacecraft system (\ref{Hill}) is actuated by a total number of thrusters $ N_\mathrm{thr} = 8 $ and chose  
$ 4 $ control directions, identified by $4$ versors $d^b_{\mathrm{thr}_i}\in \mathbb{R}^3$, $i = 1,\ldots, 4 $, which form a simplex of vectors in $ \mathbb{R}^3 $.
%
%
%

The $ 8 $ actuators are organized in $ 4 $ pairs, \cite{capello2017simplex}, and the thrusters of the $i^{th}$ pair exert their mono-directional thrusts along the same direction. 
%
To ensure that the nominal moment due to the $i^{th}$ pair of thrusters is zero, the forces exerted by the two actuators of the $i^{th}$ pair are chosen of the same magnitude 
(i.e. $T_{\mathrm{max}_i}=T_{\mathrm{max}_{i+4}}$, $i = 1,\ldots, 4 $). 
The nominal force applied to the Chaser by the $i^{th}$ pair of thrusters 
(i.e. by thruster $i^{th}$ and thruster ${(i+4)}^{th}$ together) is 
\beqs
F^b_{\mathrm{thr}_i} + F^b_{\mathrm{thr}_{i+4}} = \beta_i n T_{\mathrm{max}_i} d^b_{\mathrm{thr}_i} ,
\eeqs
$i = 1,\ldots, 4 $ and $n=2$ reflects the fact that the two thrusters are switched on simultaneously; $ \beta_i = 0 $ ($ \beta_i = 1 $), $i = 1,\ldots, 4 $, when the $i^{th}$ pair of actuators is off (on).
%


By definition the vectors $d^b_{\mathrm{thr}_i}\in \mathbb{R}^3$, $i = 1,\ldots, 4 $, form a simplex of vectors in $ \mathbb{R}^3 $, considering the rotation matrix $ R_{LVLH}(\phi,\theta,\psi)$ in (\ref{Fthr}), we can obtain a further simplex of vectors in $ \mathbb{R}^3 $ as follows
\beq
\label{simplex_d}
d_{\mathrm{thr}_i} = R_{LVLH}(\phi,\theta,\psi) d^b_{\mathrm{thr}_i} , \; i = 1,\ldots, 4.
\eeq
As a result, according to (\ref{eqt:Qj}), the space $ \mathbb{R}^3 $ is partitioned in $4$ cones by the simplex of vectors (\ref{simplex_d})
\begin{equation}
\label{eqt:Qj_d}
{Q}_j = \mathrm{cone} \left( d_{\mathrm{thr}_i}: \; i = 1, \ldots , {4}, \; i \neq j \right) .
\end{equation}
When the $i^{th}$ pair of thrusters, $i = 1,\ldots, 4 $, is switched on, it is generated the corresponding control structure
%
\begin{equation}
\label{eqt:Fthr}
{F}_{\mathrm{thr}_i} = \beta_i n T_{\mathrm{max}_i} d_{\mathrm{thr}_i}, \; i = 1,\ldots, 4 ,
\end{equation}

with $T_{\mathrm{max}_i}$ constant for all the maneuver.

The moving simplex SMC, \cite{BarPunZol11, BarPun15}, for the position tracking of system (\ref{Hill}) with sliding output $ \sigma_x $ (\ref{surf_x}) is defined by the discontinuous switching logic
\beq
\label{tau}
\begin{array}{l}
	{\rm if} \qquad
	\sigma_x \in Q_h
	\qquad
	{\rm then}             \qquad \nonumber \
	{F}_{\mathrm{thr}} = {F}_{\mathrm{thr}_h} ,
\end{array}
\eeq
where $ h \in \{1,\ldots, 4 \}$ is the least possible index, $ Q_h $ is defined by (\ref{eqt:Qj_d}) and $ {F}_{\mathrm{thr}_h} $ is given by (\ref{eqt:Fthr}).

\subsection{Spacecraft Position Control: Component-wise SMC}
\label{sec:Component-wise}
As for the simplex-based SMC, the sliding output $\sigma_x \in \mathbb{R}^3$ is defined as Equation (\ref{surf_x}).
The input vector ${F}_{\mathrm{thr}} \in \mathbb{R}^3$ can be designed according to the following first-order component-wise sliding mode control strategy
\begin{equation*} \label{tau_component-wise}
{F}_{\mathrm{thr}} = -{\it K}_0 \sgn(\sigma_x) ,
\end{equation*}
with $K_0 = nT_{max}$ and $n= 2$. For the component-wise case, the thruster configuration is the same proposed in~\cite{capello2017sliding} with 12 thrusters, two along each principal axis and for each direction (positive and negative). The Thrusters Switching Algorithm is reported in \cite{capello2017sliding}.

\section{Simulation Results}
\label{sec:Simulation Results}
Two phases of the rendezvous maneuver are analyzed, as in \cite{fehse2003automated}: (i) a \textit{radial boost}, in which two moving obstacles are introduced, and (ii) a final approach phase, to be performed in a \textit{cone} safety area. 
The Chaser has a cubic-shape of $1.2~m$ and its initial mass is $600~kg$, that decreases due to the fuel consumption.
The simulation starts $3~km$ far from the Target along V-bar. For the radial boost the goal is fixed at $[-200,0,0]~m$ and the final approach ends few centimeters before reaching the Target, to avoid docking coupling dynamics. 
The component-wise and the simplex-based SMC strategies are compared for the position tracking and, as already introduced before.
Two different thruster configurations are considered: (i) $8$ thrusters of a maximum thrust of $1.5$ N for the simplex-based SMC and (ii) $12$ thrusters of a maximum thrust of $1$ N for the component-wise SMC. The same total thrust is provided to the system by the two configurations, i.e. the sum of the thrusts provided by the actuators in each configuration is the same.
Different switching frequencies are considered, to avoid high consumption and to be compliant with the \textit{on-board} hardware constraints: 
\begin{enumerate}
    \item an update frequency of $1$ Hz for the APF algorithm and of $10$ Hz for the SMC strategies are considered for the radial boost.
    \item an update frequency of $10$ Hz for the APF guidance and a frequency of $20$ Hz are imposed for the \textit{cone approach}.
\end{enumerate}

The efficiency of the proposed position controllers is evaluated by the control effort, which represents a fuel consumption estimation. It is evaluated as $CE = \sum_{k=0}^{T_{th}} \vert F_{th,k}\vert \Delta t$, where $T_{th}$ is the switch-on time of thrusters (i.e. when the thrusters are switched on), $\vert F_{th,k}\vert = |F_x|+|F_y|+|F_z|$ and $\Delta t$ is the sample time of the simulation. 
The simulations were performed taking into account the forces and moments due to the external disturbances, as briefly described in Subsection \ref{Spacecraft Dynamics}. In particular, $\Delta F_{ex}$ includes: i) the constant aerodynamic drag of about $10^{-4}~N$ only along V-bar, ii) the force due to the \textit{J2} effect considered as a random value of about $10^{-3}~N$, iii) the constant force due to the solar pressure of about $10^{-5}~N$ along the three axes.

\begin{figure}[t]
\centering
\vspace{3mm}
\includegraphics[width=0.45\textwidth]{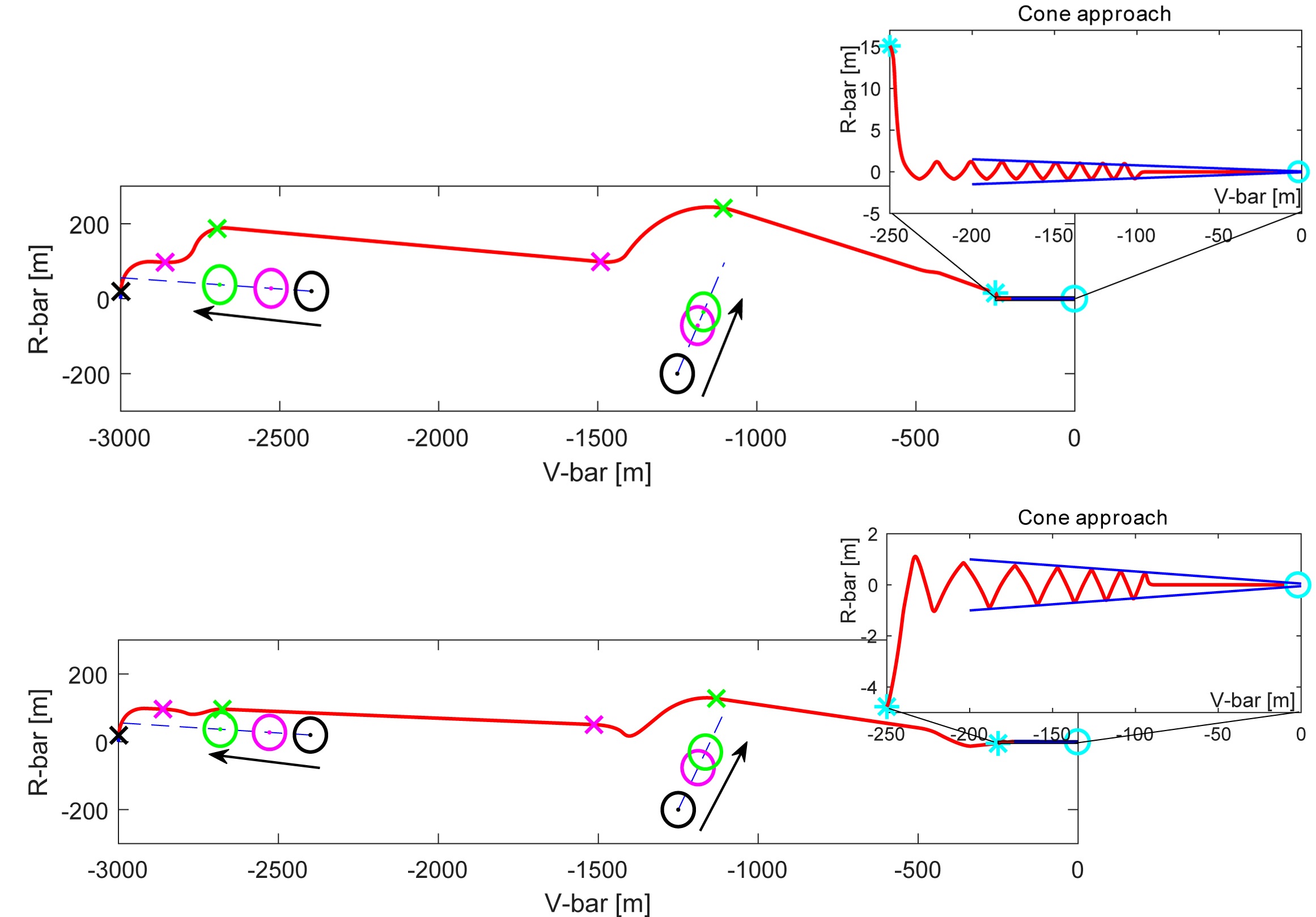}
\caption{Complete trajectory in a V-bar and R-bar plane. Upper part: Component-wise SMC; Lower Part: Simplex SMC}
\label{fig:RV trajectories}
\end{figure}

Figure~\ref{fig:RV trajectories} shows the whole rendezvous trajectories in the V-bar and R-bar plane, with a zoom for the cone approach maneuver.
The black marker \textbf{x} on the trajectory represents the starting point of the radial boost.
The arrows show the direction in which the obstacles are moving, while the circles identify the radius of the obstacle. 
The obstacles are represented with three different colors: (i) black color for the initial conditions, (ii) magenta color indicates where the sensor detects the obstacle, and (iii) green color indicates when the obstacle is overcame. These two last colors represent the evolution points of the moving obstacles.

The trajectory in V-bar and H-bar plane is omitted since the $y$ coordinate is tracked to the zero value, and in the final point $y = 10^{-5}~m$, to avoid out of plane X-Z movements.

\subsection{Radial Boost}
Even if in this phase the final goal is to reach $x = [-200,0,0]~m$, the Radial boost can be completed $50~m$ (along V-bar) before  reaching the final goal (the cyan marker in the Figure \ref{fig:RV trajectories}).
When the final point is reached, the Chaser speed is decreased to enter in the \textit{cone} safe area and to stay as close as possible to the Target position. As in Figure \ref{fig:RV trajectories}, the Chaser is able to avoid obstacles with a smooth trajectory, guaranteeing a safe maneuver. 
\begin{figure}
\centering
\hspace{-5mm}
\includegraphics[width=0.5\textwidth]{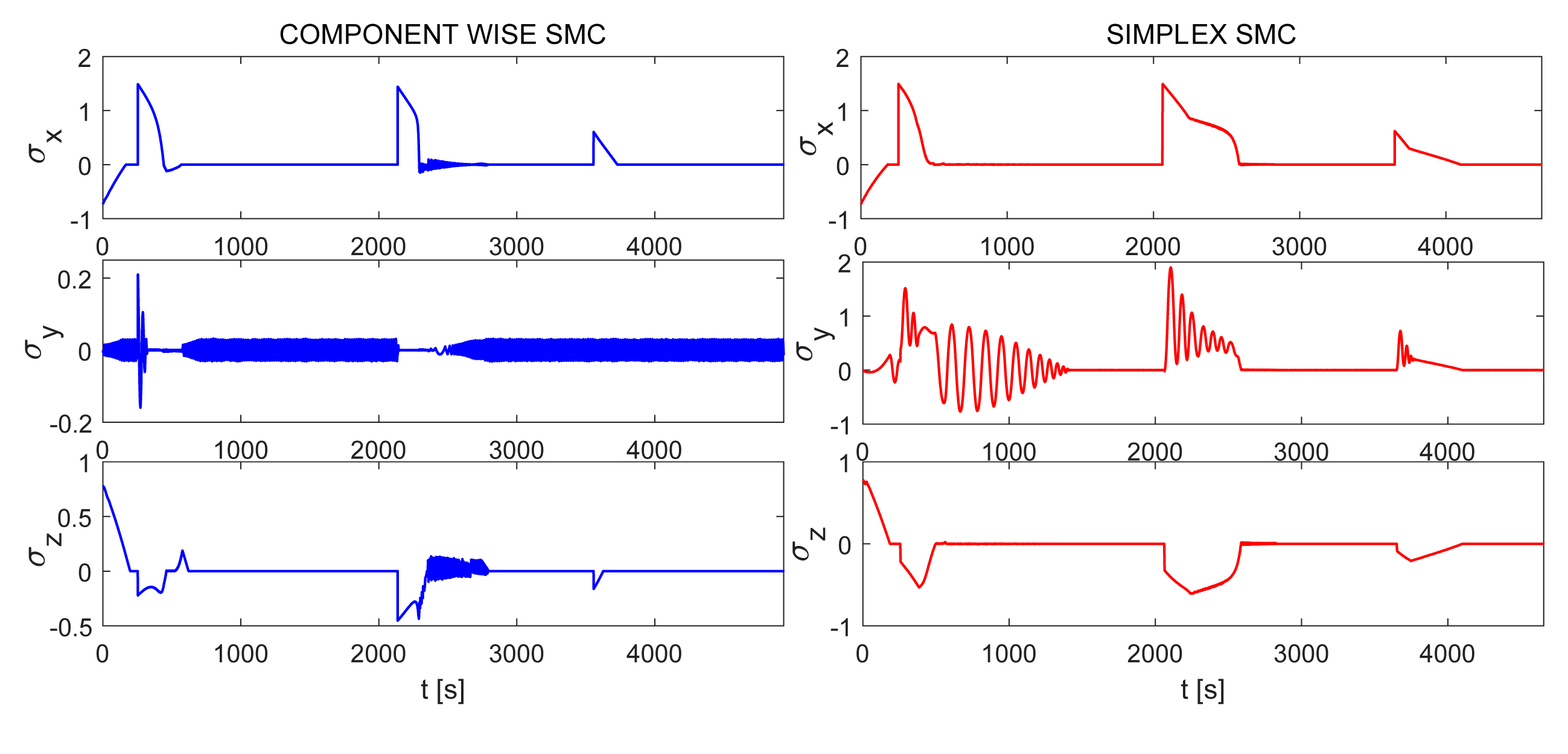}
\caption{Sliding variables for the position dynamics for radial boost}
\label{fig:sigma_pos_RB}
\end{figure}
The sliding outputs of both SMC strategies are in Figure \ref{fig:sigma_pos_RB}. It can be observed a residual chattering of an order of magnitude of $10^{-4}$. These oscillations can be filtered in real applications and can be related to the actuator noise.
The performance indicators are summarized in Table \ref{results_rb}. The fuel consumption is almost halved with a simplex-based approach, even if the maximum thrust considered is the same, as previously explained. Moreover, the control effort is strongly reduced, even if the simulation time for both cases is similar. This behavior is due to the switching logic of the simplex-based SMC: less commutation frequency is required to perform the desired maneuver. 
Note that in Table \ref{results_rb} CASE A is related to the component-wise SMC and CASE B is related to the simplex-based strategy.
\\ 
\begin{table}\vspace{3mm}
\centering
\caption{Performance Indicators for the Radial Boost}
\label{results_rb}
\begin{tabular}{ccccc}
      \hline\hline
Case & Fuel consumption & Total time & Control Effort\\
\hline
CASE A & $m_f =11.4$ kg & $t_{tot} \cong 4989$ s & $CE = 24690$ Ns\\
CASE B & $m_f =6.48$ kg & $t_{tot} \cong 4656$ s& $CE = 13968$ Ns\\
\hline\hline
\end{tabular}
\end{table}
\subsection{Cone Approach}
Due to the strict requirements in the proximity phase, no obstacles are considered. 
In Figure \ref{fig:RV trajectories}, we can observe that the Chaser is reaching the Target not exceeding the cone area. Strict requirements are required for the final approach. A final constraint on $R_{bar}$ is considered to assure the docking between the two spacecraft. The maximum value along $R_{bar}$ is $z_{f,max} = 0.05$ m. The following results are obtained: (A) $z_f =  6.9\cdot 10^{-6}$ m for the component-wise SMC and (B) $z_f =  1.8\cdot 10^{-6}$ m for the simplex-based SMC.



\begin{figure}
\centering
\vspace{3mm}
\includegraphics[width=0.4\textwidth]{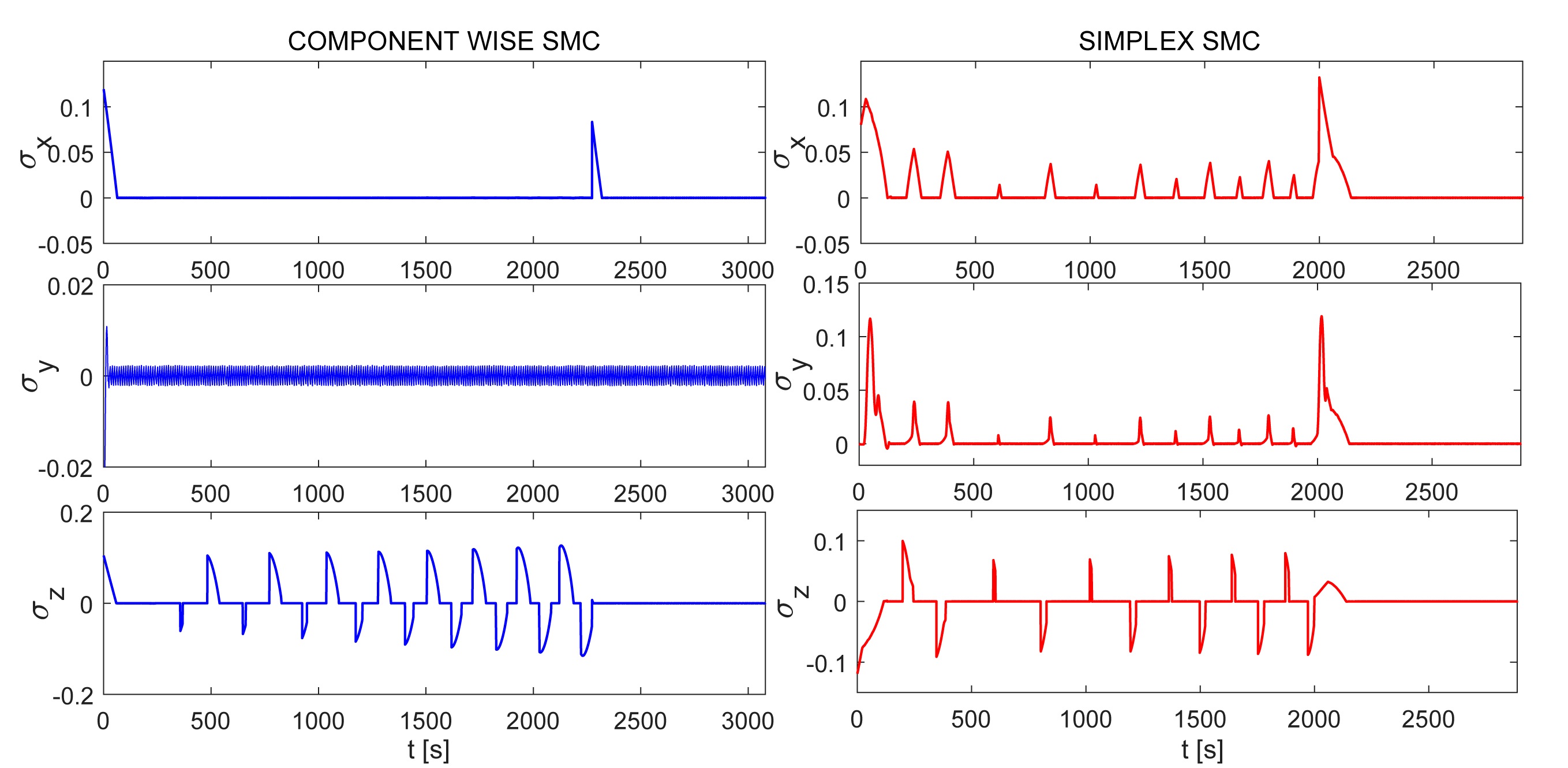}
\caption{Sliding variables of the position tracking for the cone approach}
\label{fig:sigma_pos_C}
\end{figure}

The sliding surfaces of both strategies steer to zero (Figure \ref{fig:sigma_pos_C}) with a residual error of $10^{-4}$.
The performance indicators are summarized in Table \ref{results_cone}. The fuel consumption is similar for both cases, but the control effort and the simulation time are reduced for the simplex-based SMC.
As before, note that in Table \ref{results_rb} CASE A is related to the component-wise SMC and CASE B is related to the simplex-based strategy.

\begin{table}
\centering
\caption{Performance Indicators for the cone approach maneuver} \label{results_cone}
\begin{tabular}{cccc}
      \hline\hline
      \small
Case  & Fuel consumption & Total time & Control Effort\\
\hline
CASE A &  $m_f =4.04$ kg & $t_{tot} \cong 3079$ s & $CE = 8716$ Ns\\
CASE B & $m_f =4.01$ kg & $t_{tot} \cong 2886$ s& $CE = 8658$ Ns\\
\hline\hline
\end{tabular}
\end{table}

\section{Conclusions}
\label{sec:Conclusions}

In this paper a guidance algorithm based on APF is proposed for a dynamic collision avoidance in the rendezvous maneuver. This method is combined with SMC techniques for position and attitude control of the spacecraft. As for the position control two different first order SMC methods are used. The efficiency of the two approaches is compared. 
The results shows that the simplex strategy has a better performance in terms of time, fuel consumption, control effort and chattering reduction.

\section*{Appendix A: External Orbital Disturbances}

In this Appendix, external orbital disturbances are detailed, to better clarify the mathematical model of Section \ref{sec:Mathematical Models}.

In particular we would to describe in detail how these disturbances were estimated and included in the simulations.
According with our case study, in which we analyze a rendezvous maneuver in a Low Earth Orbit (LEO), we consider four main kind of disturbances affecting the spacecraft dynamics: \textit{i}) aerodynamic drag, \textit{ii}) J2 effect, \textit{iii}) solar radiation pressure and~\textit{iv}) gravitational effect~\cite{markley2014fundamentals,zagorski2012modeling}. 

\begin{figure}[h]
\centering
\includegraphics[width=0.5\textwidth]{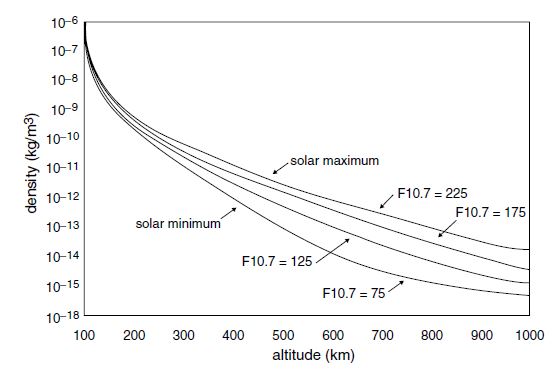}
\caption{Air density in function of the altitude}
\label{fig:aird}
\end{figure}

\begin{itemize}
    \item The \textbf {aerodynamic drag} is due to the residual air particles in space environment. Actually, Figure \ref{fig:aird} shows that the air density at the altitude of this case study, i.e. $500~km$, is very low. However the collisions occur with a great relative speed, so this effect can not be disregarded. In this work we evaluated it using Equation \ref{eq:F_AD}:
     \begin{equation}
         F_{AD}=\frac{1}{2}\rho C_D A V^2
         \label{eq:F_AD}
     \end{equation}
     The terms in the equation are estimated basing on both the altitude of the orbit and the size of the spacecraft as it follows:
     \begin{itemize}
         \item the air density $\rho$ is set at the value of $10^{-12}~kg/m^3$, according to Figure~\ref{fig:aird},
         \item $C_D=2.2$ is the drag coefficient for compact satellite,
         \item $A$ is the cross section of the satellite, i.e. $l_y\cdot l_z = 1.2^2~m^2$,
         \item $V$ is the orbital speed, i.e. $V=\omega r$, where $r=6878~km $ is the radius of the orbit and  $\omega=\sqrt{\frac{\mu}{r^3}}=1.11\cdot 10^{-3}~rad/s$ is the orbital angular speed. In the previous equation, $\mu=3.986\cdot 10^{14}$ is the Earth gravitational parameter. Then, the orbital speed value is $V=7613~km/s$.
     \end{itemize}
     Finally, evaluating the Equation \ref{eq:F_AD}, the aerodynamic drag exerts a force $F_{AD}=9.18\cdot 10^{-5}~N$ directed opposite to orbital speed, i.e. along \textit{-Vbar}.
     
     \item The \textbf{J2 effect} is due to the oblateness of the Earth. The force produced by this effect evaluated by using Equation \ref{eq:J2}, taken by \cite{sidi1997spacecraft}.
     \begin{equation}\begin{split}
        F_{J2}=-m_c\frac{3J_2\mu R_E^2}{2r^4}\left[\begin{matrix}1-3\sin^2i~\sin\theta \\ 2\sin^2i~\sin\theta~\cos\theta \\ 2\sin i~\cos i~\sin\theta \end{matrix}\right]
         \label{eq:J2}
     \end{split}\end{equation}
     Where:
     \begin{itemize}
     \item $m_c=600~kg$ is the mass of the Chaser,
     \item $J_2=1.08263\cdot 10^{-6}$ is a constant value,
     \item $\mu=3.986\cdot 10^{14}$ is the Earth gravitational parameter,
     \item $R_E=6378~km$ is the Earth radius,
     \item $r=6878~km$ is the radius of the orbit,
     \item $i$ is the orbit inclination,
     \item $\theta$ is the true anomaly.
     \end{itemize}
     So, in our case study $m_c\frac{3J_2\mu R_E^2}{2r^4}\simeq 10^{-3}~N$, while the terms into the square parenthesis are not taken into account since the inclination of the orbit is not specified, furthermore these terms do not affect the order of magnitude of this orbital disturbance. Then, this perturbation is evaluated as a force of the order of  $10^{-3}~N$, acting along the three axes. Within the simulation we introduce this force as a random value of the order of $10^{-3}~N$, to consider the variation of terms within the square brackets during the orbit.

      \item The \textbf{Solar radiation pressure} produces forces and moments when the spacecraft is in the sunlight. This is produced by photons emitted by the sun which exchange momentum with the surface of the spacecraft and its magnitude depends by the activity of the Sun, due to seasonal variations according to the Sun's cycles. The force is evaluated as follows\ref{eq:F_s}: 
      \begin{equation}
       F_{s}=(1+K)p_{s}A_{\perp} 
      \label{eq:F_s}
      \end{equation}
      where:
      
      \begin{itemize}
      
      \item $K = 0 \div 1$ is the spacecraft surface reflectivity,

      \item $p_{s} = I_{s}/c$ is the solar radiation pressure at the Earth distance from the Sun, where $I_{s}=1370~W/m^{2}$ is the solar constant and $c\simeq3 \cdot 10^{8}~m/s$ is the light speed. So, $p_{s}$ is order of~$10^{-5}~W/ms$.
     
      \item $A_{\perp}$ is the spacecraft projected area normal to sun vector. Since the Chaser has a cubic-shape of $1.2^3~m^{3}$, this term is estimated of the order of $1~m^{3}$.
      \end{itemize}  
      
      Finally, $F_{s}\simeq 10^{-5}N$. 
      Whereas, the moment due to the solar radiation pressure can be evaluated as $T_{s} = r_{s} \times F_{s}$, where $r_{s}$ is the vector from body center of mass to spacecraft optical center of pressure. Since the Chaser has a cubic-shape of $1.2~m^{3}$, this term can be at most of the order of $1~m$. Therefore, $T_{s}$ is evaluated as $10^{-5} Nm$.

    \item The \textbf{gravity effect} between the Earth and spacecraft is undoubtedly the dominant disturbances. 
    The torque due to the gravity effects can be evaluated as:
    \begin{equation}
        \Vec{M_g}=3n^2\Vec{r}\times [I]\Vec{r}
        \label{eq:T_g}
    \end{equation}
     In Equation \ref{eq:T_g} there are the following terms:
     \begin{itemize}
         \item $n^2=\frac{\mu}{a^3}=1.22\cdot 10^{-6}~rad^2/s^2$ is the orbital rate, in this equation $a=6878\cdot10^3~Km$ is the major semi-axes of the orbit;
         \item $\Vec{r}$ is the unit vector from planet to spacecraft;
         \item $[I]$ is the spacecraft inertia matrix. Since the Chaser is a symmetric cubic-shaped, it is evaluated as:
         \begin{equation*}
         \begin{split}
         I=\left[\begin{matrix}J_x & 0 & 0 \\ 0 & J_y & 0 \\ 0 & 0 & J_z\end{matrix}\right]
         \end{split}\end{equation*}
         where $J_i=m_c(2l_i)^2/12=144~Kg\cdot m^2\quad i=x,y,z$ and $l_i=1.2~m$ is the side of the Chaser.\\
         So, the moment due to the gravity effects has the same value in the three body axis and it is of the order of $10^{-6} \cdot 10^{2} = 10^{-4}~Nm$. 
         
     \end{itemize}

\end{itemize}

If we are considering, for example, the position dynamics of our system, as previously described in Section \ref{sec:Mathematical Models}, these equations are written with respect to the origin of a LVLH frame, centered on the Target satellite~\cite{fehse2003automated}. 
So, the position dynamics can be expressed as
\begin{equation}
\small
\label{Hill}
\begin{cases}
\begin{aligned}
\ddot{ x} &= \frac{F_x}{m_c} + 2\omega_0\dot{z}, \\
\ddot{y} &= \frac{F_y}{m_c} -\omega^2_0 y, \\
\ddot{z} &= \frac{F_z}{m_c} - 2\omega_0\dot{ x}+3\omega_0^2 z ,
\end{aligned} 
\end{cases}
\end{equation}

where $\ddot{x}$, $\ddot{y}$ and $\ddot{z}$ are accelerations, $\dot{x}$, $\dot{y}$ and $\dot{z}$ are velocities, $x$, $y$, and $z$ are Hill positions, all the variables are written in LVLH frame. 
$F = [F_x, F_y, F_z]^T \in \mathbb{R}^3$ is the total force acting on the spacecraft and it includes both forces due to the thrusters system and external disturbances, so $F = F_{\mathrm{thr}}+\Delta F_{ex} \in \mathbb{R}^3$, in which $F_{\mathrm{thr}} \in \mathbb{R}^3$ is the force provided by the thrusters (i.e. by the control system) and $\Delta F_{ex} \in \mathbb{R}^3$ is due to the external disturbances.

As detailed in Section \ref{sec:Mathematical Models}, $\Delta F_{ex} = F_{AD}+ F_{J2}+F_{s}$. Then, $\Delta F_{ex}$ is added to the force exerted by the thrusters $F_{\mathrm{thr}}$. Finally, the total force $F = F_{\mathrm{thr}}+\Delta F_{ex}$ is rotated by means a rotation matrix from body frame to LVLH frame in order to insert it in the Hill equations for the position dynamic.\\

In the same way, even if in this paper the main focus is on position dynamics, if attitude dynamics is taken into account, Euler equations and quaternion dynamics (see Chapter 3 and 6 of \cite{markley2014fundamentals}) are modeled to analyze a complete 6 Degree Of Freedom spacecraft dynamics. The total torque applied to the Chaser consists of three terms: (a) the moment of the thrusters $M_{thr}$, (b) the torque provided by the reaction wheels (actuators used for the attitude control) $M_{RW}$ and conceived as a Super Twisting (STW) Sliding Mode Control algorithm, 
and (c) the moment due to external disturbances $\Delta M_{ex}$.
So, we have that $M_B$ is the total torque applied to the Chaser in body frame and it consists of three terms $M_B=M_{thr}+\Delta M_{ex}+M_{RW}$. 
In particular, the moment derived by the external disturbances can be derived as follows: $M_{ex} = M_{s}+M_g$. 

\addtolength{\textheight}{-12cm}   


\bibliographystyle{IEEEtran}
\bibliography{main}

\end{document}